\documentclass[prd,aps,twocolumn,nofootinbib,preprintnumbers,superscriptaddress,balancelastpage,longbibliography]{revtex4-2}

\usepackage{graphicx}
\usepackage{bm}
\usepackage{hyperref}
\usepackage{slashed}
\usepackage{aas_macros}
\usepackage{float}
\usepackage{lipsum}
\usepackage{subfigure}
\usepackage{multirow}
\usepackage{amsmath}
\usepackage{array} 
\usepackage{varwidth} 
\usepackage{tabularx}
\usepackage{braket}
\usepackage[dvipsnames]{xcolor}
\usepackage{wasysym}

\definecolor{crimson}{RGB}{220,20,60}
\newcommand{\kin}[1]{\textcolor{crimson}{#1}}   

\usepackage[normalem]{ulem}

\def\lsim{\mathrel{\raise.3ex\hbox{$<$\kern-.75em\lower1ex\hbox{$\sim$}}}}
\def\gsim{\mathrel{\raise.3ex\hbox{$>$\kern-.75em\lower1ex\hbox{$\sim$}}}}

\newcommand{\be}{\begin{equation}}
\newcommand{\ee}{\end{equation}}
\newcommand{\bea}{\begin{eqnarray}}
\newcommand{\eea}{\end{eqnarray}}

\hypersetup{
     colorlinks   = true,
     citecolor    = Green,
     urlcolor     = Green,
     linkcolor    = Green
}

\usepackage{lettrine}

\input Zallman.fd

\usepackage{listings}
\usepackage{color,xcolor}

\begin{document}

\title{Solar Neutrino Constraints on Inelastic Dark Matter Scattering \\in Light of Recent LUX-ZEPLIN Observations}

\author{Thong T.Q. Nguyen}
\thanks{{\scriptsize Email}: \href{mailto:thong.nguyen@fysik.su.se}{thong.nguyen@fysik.su.se}; \href{https://orcid.org/0000-0002-8460-0219}{0000-0002-8460-0219}}
\affiliation{Stockholm University and The Oskar Klein Centre for Cosmoparticle Physics, Alba Nova, 10691 Stockholm, Sweden}

\author{Tim Linden}
\thanks{{\scriptsize Email}: \href{mailto:linden@fysik.su.se}{linden@fysik.su.se};  \href{http://orcid.org/0000-0001-9888-0971}{0000-0001-9888-0971}}
\affiliation{Stockholm University and The Oskar Klein Centre for Cosmoparticle Physics, Alba Nova, 10691 Stockholm, Sweden}

\author{Dan Hooper}
\thanks{{\scriptsize Email}: \href{mailto:dwhooper@wisc.edu}{dwhooper@wisc.edu};  \href{https://orcid.org/0000-0001-8837-4127}{0000-0001-8837-4127}}
\affiliation{Department of Physics, Wisconsin IceCube Particle Astrophysics Center, University of Wisconsin, Madison, WI 53706, USA}

\begin{abstract}

The LUX-ZEPLIN (LZ) Collaboration recently reported the detection of a single nuclear recoil candidate event with a very high recoil energy. The lack of any corresponding low-energy events motivates models in which dark matter scattering with nuclei has a nontrivial momentum dependence or proceeds inelastically, suppressing the rate of low-energy recoils. In this study, we consider the constraints on inelastic dark matter, including scenarios favored by the LZ event, from the absence of an excess of high-energy neutrinos from the Sun in IceCube observations. We confirm the results of Pospelov \& Ramani and show, more generally, that the lack of an excess of high-energy neutrinos from the Sun strongly constrains the parameter space in this class of models.

\end{abstract}

\maketitle

\section{Introduction}
\label{sec:intro}

The LUX-ZEPLIN (LZ) collaboration recently reported an intriguing event in a search for dark matter (DM) interactions that extended the nuclear-recoil energy window to approximately $270\,{\rm keV}$~\cite{LZ:2026axp}. In a liquid xenon exposure of 2.84~tonne-years, LZ observed a single event, LZ230616, with properties consistent with a nuclear recoil of energy $248\pm23\,({\rm stat})\pm23\,({\rm sys})\,{\rm keV}$ in a region with a very low expected background. Across the DM models considered by LZ, the largest local significance was $3.4\sigma$, which was reduced to a global significance of $2.6\sigma$ after accounting for look-elsewhere effects.  

\begin{figure}[t]
\centering
\includegraphics[width=1\columnwidth]{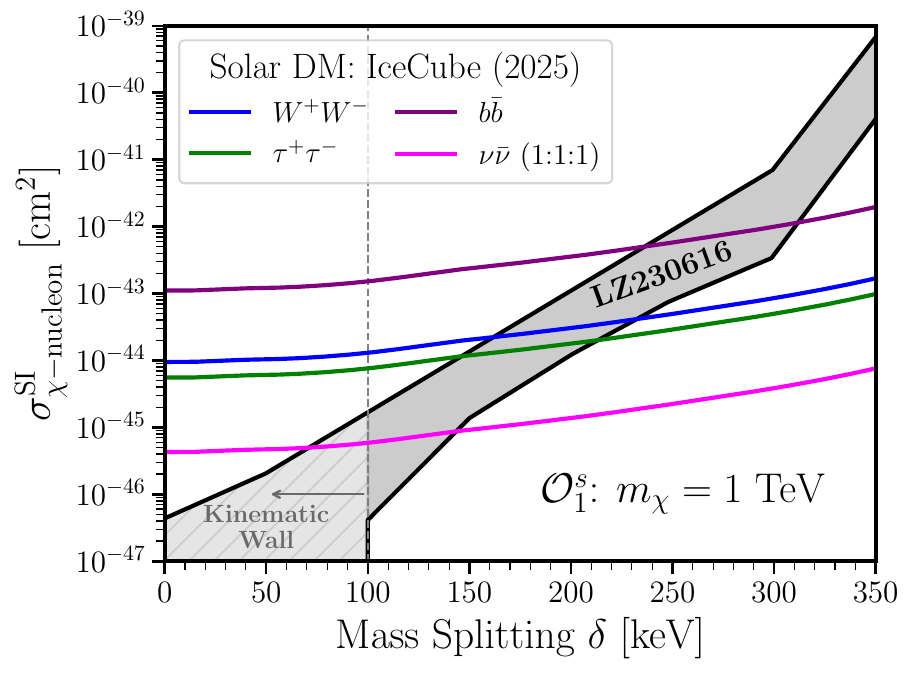}
\caption{Constraints on spin-independent inelastic scattering with isoscalar couplings ($\mathcal{O}_{1}^{s}$, corresponding to $f_p=f_n$). The parameter space that could be responsible for the LZ event is shown as a grey shaded region~\cite{LZ:2026axp}. Constraints from the absence of an excess of high-energy neutrinos from the Sun in IceCube observations~\cite{IceCube:2025fcu} are shown as solid lines, for four annihilation channels. Below the kinematic wall, adding a DM signal does not improve the fit to the LZ data.}
\vspace{-0.3cm}
\label{fig:O1s_constraints}
\end{figure}

The high recoil energy of the event reported by LZ is difficult to reconcile with conventional, momentum-independent, elastic DM scattering~\cite{Fan:2026kxx,Freese:2026sga,Wu:2026nhi,Yin:2026jnn,Su:2026rwz,Yamashita:2026ump,DiMauro:2026ldr}. The combination of the nuclear form factor and the small fraction of halo particles capable of producing such a large recoil suppresses such events relative to lower-energy recoils. Consequently, an interaction normalized to produce an observable event at this energy would generally predict many low-recoil events, in conflict with LZ data~\cite{LZ:2024zvo}. This conclusion can be avoided, however, by considering interactions with nontrivial momentum dependence~\cite{Fitzpatrick:2012ix,Anand:2013yka}. For example, pseudoscalar and mixed pseudoscalar-scalar interactions can lead to factors of $q^4$ or $q^2$ in the squared scattering amplitude, where $q$ is the momentum transfer~\cite{Berlin:2014tja,Ipek:2014gua}. Derivative interactions and electromagnetic moments can also introduce momentum and velocity dependence in the DM's scattering cross section. Such interactions can preferentially suppress the rate of low-energy recoils and enhance the relative importance of events near the upper end of the kinematically accessible recoil spectrum, making a high-energy event possible without generating too many low-energy events.

Another class of models that could explain the high-energy LZ event consists of those in which the DM particle, $\chi$, scatters with nuclei inelastically, transitioning into a heavier state, with mass splitting $\delta$~\cite{Tucker-Smith:2001myb,Tucker-Smith:2004mxa}. To leading order in $\delta/m_{\chi}$, the minimum incident speed required to produce a nuclear recoil of energy $E_R$ is 
\begin{align}
v_{\min}(E_R)=\frac{m_A E_R/\mu_{\chi A}+\delta}{\sqrt{2m_A E_R}},
\end{align}
where $m_A$ is the nuclear mass and the DM-nucleus reduced mass is $\mu_{\chi A}=m_{\chi}m_A/(m_{\chi}+m_A)$. Unlike elastic scattering, for which $v_{\min}$ increases monotonically with recoil energy, endothermic inelastic scattering has a minimum in $v_{\rm min}$ at $E_R=\mu_{\chi A} \, \delta/m_A$. Low-energy recoils therefore require increasingly large incident speeds and can become kinematically inaccessible, shifting the observable spectrum to higher energies. For sufficiently heavy DM, a splitting of a few hundred keV can place the kinematically favored recoil energy near that of the LZ event, while strongly suppressing the lower-energy signal. 

DM particles capable of producing the LZ event without generating an unacceptably large number of lower-energy events can also scatter efficiently with nuclei in the Sun~\cite{Nussinov:2009ft,Menon:2009qj}. Gravitational acceleration increases the speed of DM particles as they approach the Sun, allowing inelastic scattering on heavy nuclei even when the same process is strongly suppressed in terrestrial detectors. A collision that removes sufficient kinetic energy from a DM particle can leave it gravitationally bound. If the DM can subsequently self-annihilate, such particles may produce a flux of high-energy neutrinos that could be detected by large-volume neutrino telescopes~\cite{Press:1985ug, Gould:1987ju, Garani:2017jcj, Busoni:2017mhe}.

A very recent analysis of neutrino data by Pospelov and Ramani~\cite{Pospelov:2026ewn} utilized this mechanism to show that current IceCube data rule out the standard thermal higgsino interpretation of the LZ event, a result which has since been confirmed by Refs.~\cite{Bose:2026ndd, DiMauro:2026dqp}, the first of which also extended the analysis to include Super-Kamiokande data~\cite{Bose:2026ndd}. Intriguingly, Ref.~\cite{Pospelov:2026ewn} showed that in the case of inelastic higgsino DM, the large recoils afforded by the heaviest solar elements (such as lead and uranium) play a central role in capturing and cooling solar DM, despite their very low solar abundances. These constraints could motivate scenarios in which the higgsinos are produced non-thermally~\cite{Rodd:2026tyn}.

In this work, we extend these previous results and use IceCube observations to test a broader range of inelastic DM models, including scenarios motivated by LZ observations. In Fig.~\ref{fig:O1s_constraints}, we show that IceCube strongly constrains this parameter space, excluding the majority of inelastic DM scenarios consistent with the LZ data, so long as the DM annihilates with a significant branching fraction to a combination of neutrinos, taus, heavy quarks, or gauge and/or Higgs bosons. We also discuss DM models that could produce the LZ event while remaining consistent with neutrino observations.

\section{Inelastic Dark Matter}

Inelastic DM can arise when a pair of nearly degenerate states has an interaction that preferentially connects the two~\cite{Tucker-Smith:2001myb,Cui:2009xq}. There are many well-motivated scenarios in which this phenomenology could be realized, including the following well-known cases:

\smallskip
{\bf Pseudo-Dirac fermions coupled to a new vector mediator.} 
Consider a model with a pair of Weyl fermions, $\eta$ and $\xi$, with the following mass terms:
\begin{equation}
 -\mathcal{L}_{\rm mass}=m_D\eta\xi+\frac12m_\eta\eta\eta+\frac12m_\xi\xi\xi+\mathrm{h.c.}
\end{equation}
For real mass parameters with $m_D >0$ and $|m_D| \gg |m_{\eta}|, |m_{\xi}|$, two nearly degenerate mass eigenstates will result, $m_{\chi_{1,2}} \approx  m_D \mp |m_{\eta}+m_{\xi}|/2$. Furthermore, if the original Dirac fermion has a vector coupling to a mediator, $V_{\mu}$, the leading interaction couples off-diagonally to this pair of states, 
\begin{equation}
 \mathcal{L}_{\rm int}\supset i g_D V_\mu\bar\chi_2\gamma^\mu\chi_1.
\end{equation}
In the context of scattering with nuclei, this vector mediator could be a dark photon kinetically mixed with hypercharge, or another $Z'$ that couples to quarks.

\smallskip
{\bf Split complex scalars coupled to a new vector mediator.} For a complex scalar, $\phi$, consider 
\begin{align}
V \supset m^2_0 |\phi|^2 + \frac{1}{2}(b \phi^2 + {\rm h.c.}).
\end{align}
For $m_0^2 \gg|b|$, this leads to a pair of physical states with a small mass splitting, $m_{\phi_1, \phi_2} \approx m_0 \mp |b|/2m_0$. Again, the vector gauge interaction experienced by these two states is off-diagonal,
\begin{align}
\mathcal{L}_{\rm int} \supset g_D V_{\mu} (\phi_2 \partial^{\mu} \phi_1 - \phi_1 \partial^{\mu}\phi_2).
\end{align}

\smallskip
{\bf Electroweak multiplets.} If the DM is part of an electroweak multiplet, it is possible that inelastic scattering could be mediated by the Standard Model $Z$. Well known examples of this include higgsino-like fermions~\cite{Martin:1997ns, Wells:2003tf, Hall:2011jd, Giudice:2004tc, Arvanitaki:2012ps, Arkani-Hamed:2004ymt, Arkani-Hamed:2012fhg, Nagata:2014wma,Co:2021ion} and inert scalar doublets~\cite{Arina:2009um}. In an inert doublet scenario, the operator $\lambda_5(H^\dagger\Phi)^2/2+\mathrm{h.c.}$ splits the two neutral, real scalar components, giving rise to a mass splitting of $\delta \approx |\lambda_5|v^2/(2m_\chi)$, where $v=246\,\mathrm{GeV}$ is the Higgs vev. Higgsinos consist of two fermionic electroweak doublets with hypercharge $Y=\pm 1/2$. Their components form a charged Dirac fermion and, when higgsino-number-violating interactions are included, a pair of neutral Majorana fermions. In the context of supersymmetry with heavy electroweak gauginos and real mass parameters, the two Majorana states are nearly degenerate, with 
\begin{equation}
 \delta\approx m_Z^2\left|\frac{\sin^2\theta_W}{M_1}
 +\frac{\cos^2\theta_W}{M_2}\right| \sim \frac{m_Z^2}{M_{1,2}},
\end{equation}
where $M_1$ and $M_2$ are the bino and wino mass parameters. For a common gaugino mass of $M_{1,2} \sim 3 \times 10^7 \, {\rm GeV}$, for example, this yields a mass splitting of $\delta \sim 300 \, {\rm keV}$. A stabilizing $Z_2$ symmetry (such as $R$-parity) can render the lightest of these states stable. Under standard cosmological assumptions, the thermal relic abundance of higgsino DM is predicted to be equal to the measured DM density for a mass of $m_{\chi} \simeq 1.08 \, {\rm TeV}$. Importantly, the two higgsino states have an off-diagonal coupling to the $Z$ boson that induces spin-independent inelastic scattering, $\chi_1 N\rightarrow\chi_2 N$. Neglecting the form factor, the cross section for this process is
\begin{align}
\sigma_{\chi_1 N}(v) = \frac{G_F^2 \mu_{\chi A}^2}{2\pi} &\bigg[(A-Z)-(1-4 \sin^2 \theta_W) Z\bigg]^2 \nonumber \\
&\times\sqrt{1-\frac{2\delta}{\mu_{\chi A} v^2}},
\end{align}
where $G_F$ is the Fermi constant, $\mu_{\chi A}$ is the reduced mass of the system, $A$ is the mass number, $Z$ is the atomic number, and $v$ is the velocity of the DM particle. This expression applies for $v > \sqrt{2\delta/\mu_{\chi A}}$, with zero cross section below this threshold. In the absence of kinematic suppression due to the mass splitting, this corresponds to a cross section on a neutron of $\sigma_{\chi_1 n} \approx 7 \times 10^{-39}$~cm$^2$.

In the following section, we return to the capture and annihilation of DM in the Sun, considering not only higgsinos, but a broad range of inelastic DM models.

\section{Dark Matter in the Sun}
\label{sec:capture}

\subsection{Dark Matter Capture in the Sun}

If DM particles can scatter with xenon nuclei in the LZ detector, they will also scatter with nuclei in the Sun. These scattering processes can cause DM particles traversing the Sun to lose kinetic energy and cause them to become gravitationally captured if or when their velocity falls below the Sun's local escape velocity~\cite{Press:1985ug, Gould:1987ju}. In some scenarios, with large nuclear cross sections, a DM particle can scatter many times in the Sun, increasing the probability of capture. In all of the scenarios discussed in this paper, however, the Sun is optically thin to DM, and we calculate the DM capture rate assuming that a DM particle interacts only once before being captured or escaping. In this case, the capture rate can be expressed as follows:
\begin{align}
    C_{\odot} = \sum_{A}&\int_{0}^{R_{\odot}}{\rm d}r\,4\pi r^{2}n_{A}(r)\int_{0}^{u_{\rm max}}{\rm d}u_{\chi}\frac{f_{v_{\odot}}(u_{\chi})}{u_{\chi}}\nonumber\\
    & \times \frac{\rho_{\odot}}{m_{\chi}}w^2(r)\int_{E_{R}^{\rm min}}^{E_{R}^{+}}{\rm d}E_{R} \frac{{\rm d}\sigma_{\chi A}}{{\rm d}E_{R}},
    \label{eq:Cweak}
\end{align}
where $R_{\odot}=696,340$~km is the solar radius, $m_{\chi}$ is the DM mass, and \mbox{$\rho_{\chi}=0.3$~GeV/cm$^{3}$} is the local DM density. For the density of nuclei, $n_{A}(r)$, we adopt the standard solar model (SSM) from Ref.~\cite{Scott:2014lka}. Prior to scattering, gravitational acceleration relates the local DM speed, $w$, to the asymptotic speed, $u_{\chi}$, and the local solar escape speed, $v_{e}(r)$, as $  w^2(r)=u_{\chi}^{2}+v_{e}^{2}(r)$.
We adopt the local solar escape speed profile from Ref.~\cite{Nguyen:2025ygc} and use a shifted Maxwell-Boltzmann distribution for the velocities in the halo:
\begin{align}
    f_{v_\odot}(u_\chi)
    &= \sqrt{\frac{3}{2\pi}}\frac{u_\chi}{v_\odot v_d} \\ \nonumber 
 &\times   \left[
        \exp\!\left(-\frac{3(u_\chi-v_\odot)^2}{2v_d^2}\right)
        -
        \exp\!\left(-\frac{3(u_\chi+v_\odot)^2}{2v_d^2}\right)
    \right],
    \label{eq:fvodot}
\end{align}
where $v_{\odot}=220$~km/s and $v_{d}=270$~km/s. We also adopt a cutoff for this velocity distribution at the local Galactic escape velocity, \mbox{$u_{\rm max} \approx 533$~km/s}~\cite{Piffl:2013mla}. We note that while interpretations of the LZ observations can be highly sensitive to the precise choice of the DM velocity distribution, and in particular to its extreme high-velocity tail, this is not true for solar capture. In the case of solar capture, the high velocities stem primarily from the gravitational potential of the Sun, limiting the impact of the Galactic Halo's velocity distribution.

The capture rate also depends on the DM's scattering cross section with target nuclei. For spin-independent interactions generated by the $\mathcal{O}_{1}^{s}$ and $\mathcal{O}_{1}^{v}$ operators, the differential DM-nucleus cross section is
\begin{equation}
\frac{d\sigma_{\chi A}}{dE_R}
  = \frac{m_A\,\sigma_{\chi p}^{\rm SI}}{2\,\mu_{\chi p}^{2}\,w^2}
    \left[Z
        + \frac{f_n}{f_p}\,
          (A-Z)\right]^{2} F_A^{2}(q^{2}),
\label{eq:dsigma}
\end{equation}
where $\mu_{\chi p}$ is the reduced mass of the DM-proton system, and $\sigma_{\chi p}^{\rm SI}$ is the spin-independent cross section for DM scattering with a proton, defined in the elastic, zero-momentum-transfer limit. In particular, it does not include the inelastic threshold suppression. For $F_{A}(q)$, we adopt the Helm-form factor from Refs.~\cite{Helm:1956zz, Lin:2019uvt}, with \mbox{$q=\sqrt{2 m_{A}E_{R}}$}. $Z$ and $A$ are the atomic number and mass number of the target nucleus, respectively. We consider three ratios of the neutron-to-proton coupling, corresponding to the cases of an isoscalar operator $\mathcal{O}_{1}^{s}$ with $f_{n}=f_{p}$, an isovector operator $\mathcal{O}_{1}^{v}$ with $f_{n}=-f_{p}$, and the higgsino case with $f_p=-f_n \,(1-4 \sin^2 \theta_W)$.

These coupling choices lead to very different relative scattering rates in xenon and in the Sun. Xenon is neutron rich, with $Z=54$ and $(A-Z) \simeq 77$. The Sun, by contrast, consists primarily of hydrogen and helium, with smaller abundances of heavier elements. For nuclei with equal numbers of protons and neutrons, such as $^4$He, $^{12}$C, and $^{16}$O, the isovector proton and neutron amplitudes cancel. Hydrogen is not subject to this cancellation, but its small mass makes it inaccessible to inelastic scattering with mass splittings in the range relevant to the LZ event. Capture through isovector interactions is therefore suppressed relative to the isoscalar case.

We also consider spin-dependent interactions generated by the $\mathcal{O}_{4}^{s}$ and $\mathcal{O}_{4}^{v}$ operators~\cite{Unwin:2026rdp}. Normalizing to the reference DM-proton cross section, the differential nuclear cross section can be written as
\begin{equation}
\frac{d\sigma_{\chi A}}{dE_R}
  = \frac{m_A\,\sigma_{\chi p}^{\rm SD}}{2\,\mu_{\chi p}^{2}\,w^{2}}\,
    \frac{4}{3}\,\frac{J+1}{J}
    \left|\langle S_p\rangle
        + \frac{a_n}{a_p}\,
          \langle S_n \rangle\right|^{2}
    |F_{\rm SD}(q^{2})|^{2},
\label{eq:dsigmaSD}
\end{equation}
where $\sigma_{\chi p}^{\rm SD}$ is the reference cross section in the elastic, zero-momentum-transfer limit. We adopt the spin-dependent form factor from Ref.~\cite{Garani:2017jcj}, with $F_{\rm SD}(0)=1$. Similar to the spin-independent case, the coupling ratios are $a_p/a_n=\pm 1$ for the $\mathcal{O}_{4}^{s}$ and $\mathcal{O}_{4}^{v}$ operators, respectively. Here, $J$ is the nuclear spin, and $\langle S_p\rangle$ and $\langle S_n\rangle$ are the expectation values of the total proton and neutron spin contributions in the nuclear state with maximal spin projection. We adopt these quantities from Ref.~\cite{Bednyakov:2004xq}. Spin-zero nuclei do not contribute. Hydrogen generally dominates spin-dependent capture in the elastic limit~\cite{Nguyen:2026nhe, Nguyen:2026pdr}, but heavier spin-carrying isotopes become essential once inelastic scattering on hydrogen is kinematically forbidden. When considering spin-dependent scattering, we focus in this study on the DM-proton cross section.

In the case of inelastic scattering, the kinematic limits for the recoil energy are given by
\begin{align}
    E^{\pm}_{R}=\frac{\mu^2_{\chi A} \, (w\pm w')^{2} }{2m_{A}},
\end{align}
where $w'$ is the final relative speed of the DM particle and nucleus, evaluated to leading order in $\delta/m_\chi$:
\begin{equation}
    w'=\sqrt{w^{2}-\frac{2\delta}{\mu_{\chi A}}}.
    \label{eq:wprime}
\end{equation}
In this case, the minimum recoil energy for the DM to be captured in the Sun is
\begin{equation}
    E^{\rm min}_{R} = {\rm max}\left[\, E^{-}_{R}\, , \, \frac{1}{2}m_{\chi}u_{\chi}^{2}-\delta \,\right].
\end{equation}
The contribution to Eq.~\eqref{eq:Cweak} is zero if $w^2<2\delta/\mu_{\chi A}$ or $E_R^{\rm min}>E_R^+$. 

To calculate the solar capture rate, we sum the contributions from the 45 target isotopes included in our solar model~\cite{Scott:2014lka}, spanning from hydrogen to uranium. Although hydrogen is the most abundant solar element, its small mass makes inelastic scattering inaccessible at splittings of several hundred keV. Heavier nuclei, including iron and nickel, can support much larger splittings. Near the upper end of solar kinematic sensitivity, trace elements such as lead and uranium can dominate. Our element-by-element capture rates agree well with those of Pospelov and Ramani~\cite{Pospelov:2026ewn}.

\subsection{Dark Matter Thermalization in the Sun}

Once a DM particle has become gravitationally bound to the Sun, repeated solar passages provide further opportunities to scatter. The cumulative scattering probability depends on its orbit, the target densities, and the relevant cross sections. These interactions determine the spatial and velocity distributions of the captured population and can strongly affect the resulting annihilation rate.

Immediately after capture, DM particles have a velocity distribution that is highly non-thermal. A particle captured near the Sun's core, for example, can retain a kinetic energy of order \mbox{$\tfrac{1}{2}m_{\chi}v^{2}_{\rm e}\sim10$~MeV} for $m_{\chi}=1$~TeV, compared to a mean thermal kinetic energy of $\tfrac{3}{2}T_{c} \approx 2$~keV. The DM particle must lose this remaining energy in order to thermalize with its environment. Each subsequent collision transfers a mean fraction $\sim 2\mu_{\chi A}^{2}/m_{\chi}m_{A} \sim 2m_{A}/m_{\chi}$ of the DM's kinetic energy to the recoiling nucleus, making heavy nuclei important to the process of DM thermalization.

An inelastic collision removes an additional $\delta$ from the DM's kinetic energy, but the resulting $\chi_2$ state then downscatters on a timescale much shorter than its decay time. Thus, over a complete upscattering-downscattering cycle, the internal energy changes cancel, and the net DM energy loss is the sum of the nuclear recoil energies. The recoil term therefore sets the net cooling rate in both the elastic and the inelastic case. The role of $\delta$ is only to impose a threshold below which inelastic scattering stops.

Repeated inelastic collisions quickly cool the DM particles until their velocities fall below the kinematic threshold discussed in the previous section. For a stationary target, upscattering requires $w>\sqrt{2\delta/\mu_{\chi A}}$, corresponding to a kinetic energy of $E>m_\chi\delta/\mu_{\chi A}$. As the DM cools, scattering on the heaviest available nuclei therefore becomes increasingly important. For $m_\chi=1$~TeV and $\delta=350$~keV, iron and nickel remain accessible down to $E\sim7$~MeV, whereas lead and uranium remain accessible down to $E\sim2$~MeV. The small abundances of these heaviest elements make each such collision rare, but still allow the DM to reach energies of $E \sim \delta \,  (m_{\chi}+m_A)/m_A$  on timescales shorter than the age of the Sun.

\begin{figure*}[t!]
\centering
\includegraphics[width=1\columnwidth]{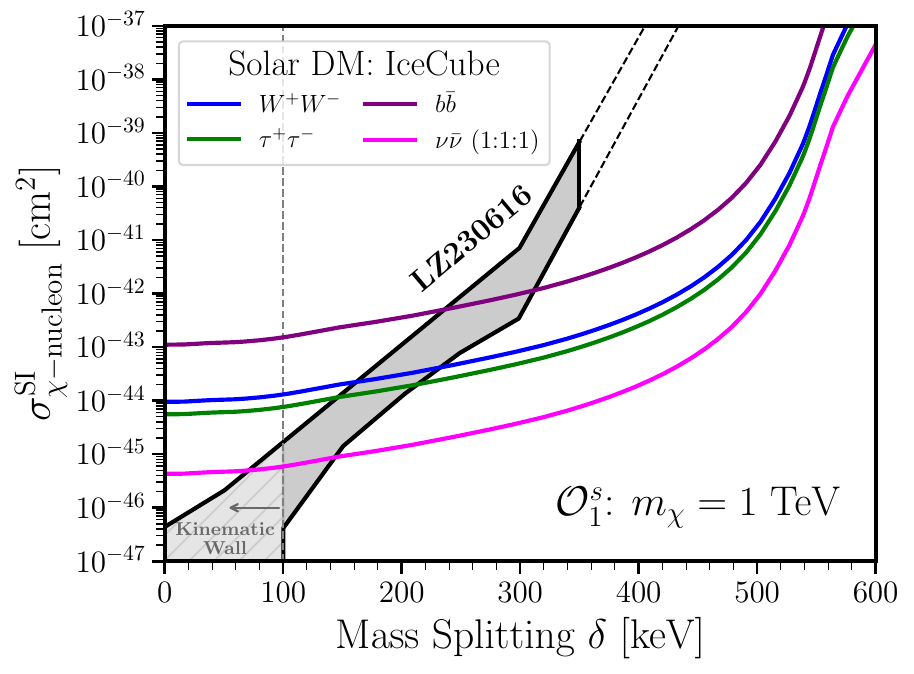}
\hfill
\includegraphics[width=1\columnwidth]{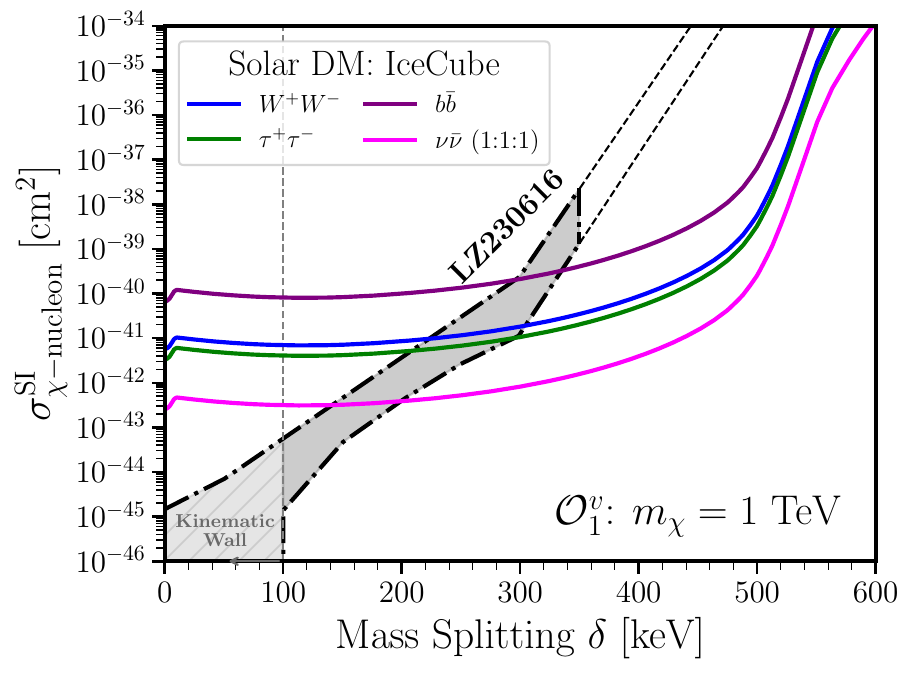}
\caption{As in Fig.~\ref{fig:O1s_constraints}, but for spin-independent, isoscalar ($\mathcal{O}^s_1$, $f_p=f_n$) or isovector ($\mathcal{O}^v_1$, $f_p=-f_n$) couplings. The dashed lines represent a highly simplified logarithmic extrapolation of the results reported by the LZ collaboration~\cite{LZ:2026axp}, and exist solely to remind the reader that the high-mass splitting limit of the LZ analysis stems from analysis choices set by the LZ collaboration. Computing such an extrapolation correctly depends sensitively on the assumed DM halo velocity distribution.}
\label{fig:O1_extend}
\end{figure*}

\begin{figure*}[t!]
\centering
\includegraphics[width=1\columnwidth]{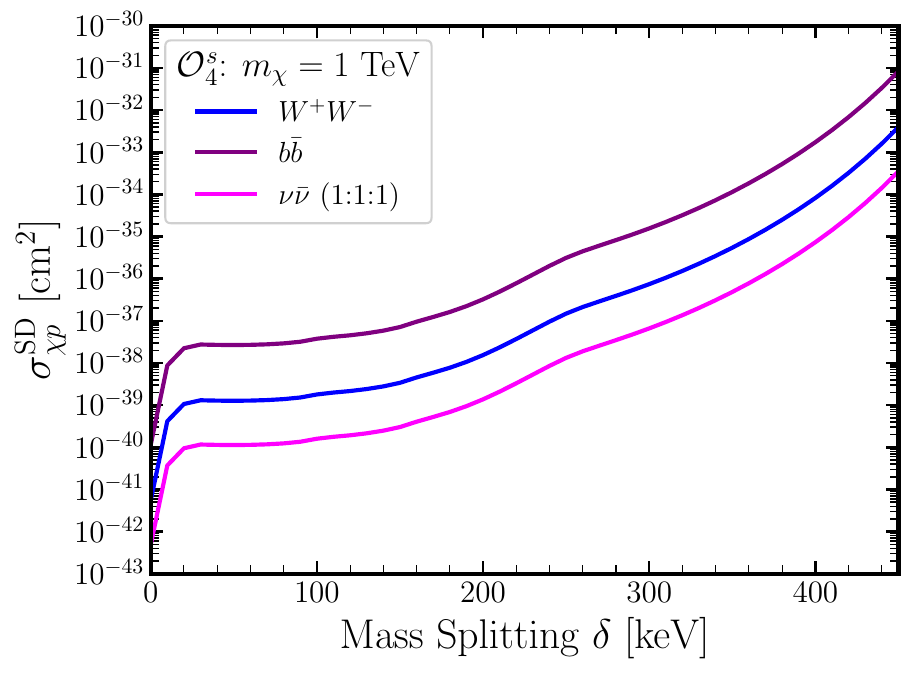}
\hfill
\includegraphics[width=1\columnwidth]{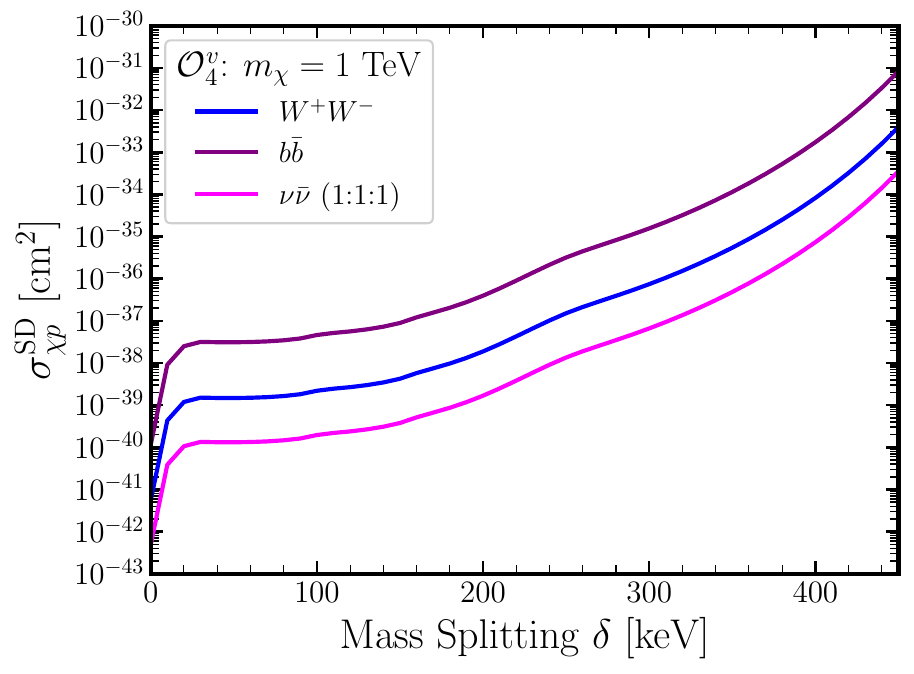}
\caption{As in Fig.~\ref{fig:O1_extend}, but for spin-dependent, isoscalar ($\mathcal{O}^s_4$, $a_p=a_n$) or isovector ($\mathcal{O}^v_4$, $a_p=-a_n$) couplings.}
\label{fig:O4_extend}
\end{figure*}

Inelastic scattering can concentrate captured DM toward the solar core without thermalizing it. In many scenarios, however, the DM will experience additional elastic interactions that continue to cool the DM after inelastic channels have closed. For higgsinos, elastic scattering occurs at the one-loop level, with $\sigma_{\rm SI} \sim 10^{-50}~\mathrm{cm}^{2}$, although cancellations introduce substantial uncertainty~\cite{Pospelov:2026ewn}. Elastic spin-dependent scattering on hydrogen provides an additional cooling channel that can allow for capture-annihilation equilibrium, even when the spin-independent contribution is strongly suppressed~\cite{Pospelov:2026ewn}. Summed over all solar targets, higgsino DM is predicted to reach kinetic equilibrium in the Sun's core after $\sim10^{16}$~s, less than the age of the Sun. Note that the total cooling time is controlled by the rate of the final collisions, since the scattering rate falls as the DM cools.

As the DM thermalizes in the Sun's core, it relaxes into an isothermal distribution of scale radius
\begin{equation}
 r_{\rm therm}=\left(\frac{3k_B T_c}{2\pi G\rho_c m_\chi}\right)^{1/2} \approx 0.0037\,R_{\odot} \, \bigg(\frac{1 \, {\rm TeV}}{m_{\chi}}\bigg)^{1/2},
\end{equation}
where $\rho_c$ is the central solar mass density. This radius is much less than the $\sim0.02\,R_{\odot}$ required for capture-annihilation equilibrium to be maintained (as discussed in the next section)~\cite{Pospelov:2026ewn}. Notably, there is one caveat to this conclusion. The loop-level higgsino spin-independent cross section may be subject to a cancellation and may fall well below its central value. Should it and all other such channels vanish, DM cooling could stall before thermalizing. More generally speaking, the thermalization of the DM requires a small but non-zero rate of elastic scattering, which may or may not be present in a given model. In what follows, we assume that the captured DM is able to reach thermal equilibrium with the Sun's core. Otherwise, our constraints could weaken considerably.

\subsection{Neutrinos from Dark Matter Annihilation in the Sun}
\label{ssec:annihilate}


\begin{table*}[p]
\centering
\renewcommand{\arraystretch}{1.4}
\resizebox{\textwidth}{!}{%
\begin{tabular}{c|c|ccccccccccc}
\hline\hline
\multirow{3}{*}{$m_\chi$} & & \multicolumn{11}{c}{IceCube 90\% CL limits on $\mathcal{O}_{1}^{s}$ $\sigma_{\rm SI}$ $[\times 10^{-45}\,\text{cm}^{2}]$} \\[2pt] \cline{3-13}
 & \rule{0pt}{12pt} SM channel & \multicolumn{11}{c}{Mass Splitting $\delta$ [keV]} \\[2pt]
 & & 100 & 150 & 200 & 250 & 300 & 350 & 400 & 450 & 500 & 550 & 600\\ \hline
\multirow{4}{*}{400 GeV}
 & $W^+W^-$      & 15  & 25  & 40  & 70  & 131 & \kin{294} & \kin{902} & \kin{$5.2\!\times\!10^{3}$} & \kin{$1.0\!\times\!10^{5}$} & \kin{$8.3\!\times\!10^{6}$}  & \kin{$1.1\!\times\!10^{10}$} \\
 & $\tau^+\tau^-$ & 15  & 25  & 39  & 68  & 127 & \kin{286} & \kin{879} & \kin{$5.1\!\times\!10^{3}$} & \kin{$9.8\!\times\!10^{4}$} & \kin{$8.1\!\times\!10^{6}$}  & \kin{$1.0\!\times\!10^{10}$} \\
 & $b\bar{b}$     & 647 & 1068 & 1694 & 2952 & 5541 & \kin{$1.2\!\times\!10^{4}$} & \kin{$3.8\!\times\!10^{4}$} & \kin{$2.2\!\times\!10^{5}$} & \kin{$4.3\!\times\!10^{6}$} & \kin{$3.5\!\times\!10^{8}$}  & \kin{$4.5\!\times\!10^{11}$} \\
 & $\nu\bar{\nu}$ & 0.60 & 0.99 & 1.6 & 2.7 & 5.1 & \kin{12} & \kin{36} & \kin{204} & \kin{$4.0\!\times\!10^{3}$} & \kin{$3.3\!\times\!10^{5}$}  & \kin{$4.2\!\times\!10^{8}$}  \\[4pt] \hline
\multirow{4}{*}{1 TeV}
 & $W^+W^-$      & 13  & 20  & 30  & 50  & 85  & 168 & \kin{425} & \kin{1653} & \kin{$1.8\!\times\!10^{4}$} & \kin{$3.3\!\times\!10^{6}$}  & \kin{$9.9\!\times\!10^{8}$}  \\
 & $\tau^+\tau^-$ & 7.6 & 12  & 18  & 29  & 50  & 98  & \kin{249} & \kin{966}  & \kin{$1.0\!\times\!10^{4}$} & \kin{$1.9\!\times\!10^{6}$}  & \kin{$5.8\!\times\!10^{8}$}  \\
 & $b\bar{b}$     & 151 & 238 & 355 & 578 & 994 & 1962 & \kin{4960} & \kin{$1.9\!\times\!10^{4}$} & \kin{$2.1\!\times\!10^{5}$} & \kin{$3.9\!\times\!10^{7}$}  & \kin{$1.1\!\times\!10^{10}$} \\
 & $\nu\bar{\nu}$ & 0.59 & 0.92 & 1.4 & 2.2 & 3.9 & 7.6 & \kin{19}  & \kin{75}   & \kin{806}  & \kin{$1.5\!\times\!10^{5}$}  & \kin{$4.5\!\times\!10^{7}$}  \\[4pt] \hline
\multirow{4}{*}{4 TeV}
 & $W^+W^-$      & 158 & 243 & 353 & 557 & 918 & 1707 & 4017 & \kin{$1.4\!\times\!10^{4}$} & \kin{$1.2\!\times\!10^{5}$} & \kin{$5.7\!\times\!10^{7}$}  & \kin{$6.4\!\times\!10^{9}$}  \\
 & $\tau^+\tau^-$ & 50  & 77  & 112 & 177 & 291 & 542  & 1276 & \kin{4439} & \kin{$3.9\!\times\!10^{4}$} & \kin{$1.8\!\times\!10^{7}$}  & \kin{$2.0\!\times\!10^{9}$}  \\
 & $b\bar{b}$     & 271 & 416 & 604 & 951 & 1568 & 2916 & 6861 & \kin{$2.4\!\times\!10^{4}$} & \kin{$2.1\!\times\!10^{5}$} & \kin{$9.7\!\times\!10^{7}$}  & \kin{$1.1\!\times\!10^{10}$} \\
 & $\nu\bar{\nu}$ & 10  & 16  & 23  & 36  & 59  & 109  & 257  & \kin{893}  & \kin{$7.9\!\times\!10^{3}$} & \kin{$3.6\!\times\!10^{6}$}  & \kin{$4.1\!\times\!10^{8}$}  \\[4pt] \hline\hline
\end{tabular}%
}
\caption{IceCube 90\% confidence level upper limits on the reference spin-independent DM-proton cross section for the isoscalar $\mathcal{O}_1^s$ interaction. Entries in red denote splittings for which a 248~keV xenon recoil is kinematically forbidden in the adopted halo model, because $v_{\min}(E_R)=(m_AE_R/\mu_{\chi A}+\delta)/\sqrt{2m_AE_R}$ exceeds the maximum incident speed at Earth. Solar scattering can remain accessible because of gravitational acceleration and heavier targets.}
\label{tab:IceCube_O1s}
\end{table*}


\begin{table*}[p]
\centering
\renewcommand{\arraystretch}{1.4}
\resizebox{\textwidth}{!}{%
\begin{tabular}{c|c|ccccccccccccc}
\hline\hline
\multirow{3}{*}{$m_\chi$} & & \multicolumn{13}{c}{IceCube 90\% CL limits on $\mathcal{O}_{1}^{v}$ $\sigma_{\rm SI}$ $[\times 10^{-40}\,\text{cm}^{2}]$} \\[2pt] \cline{3-15}
 & \rule{0pt}{12pt} SM channel & \multicolumn{13}{c}{Mass Splitting $\delta$ [keV]} \\[2pt]
 & & 0 & 50 & 100 & 150 & 200 & 250 & 300 & 350 & 400 & 450 & 500 & 550 & 600 \\ \hline
\multirow{4}{*}{400 GeV}
 & $W^+W^-$         & 0.053 & 0.086 & 0.077 & 0.083 & 0.11 & 0.16 & 0.27 & \kin{0.60} & \kin{1.8} & \kin{11.1} & \kin{917} & \kin{$8.9\!\times\!10^{5}$} & \kin{$2.3\!\times\!10^{8}$} \\
 & $\tau^+\tau^-$   & 0.052 & 0.084 & 0.075 & 0.081 & 0.10 & 0.15 & 0.27 & \kin{0.59} & \kin{1.8} & \kin{10.8} & \kin{894} & \kin{$8.7\!\times\!10^{5}$} & \kin{$2.2\!\times\!10^{8}$} \\
 & $b\bar{b}$       & 2.3 & 3.7 & 3.3 & 3.5 & 4.6 & 6.7 & 11.7 & \kin{25.5} & \kin{77.5} & \kin{469} & \kin{$3.9\!\times\!10^{4}$} & \kin{$3.8\!\times\!10^{7}$} & \kin{$9.7\!\times\!10^{9}$} \\
 & $\nu\bar{\nu}$   & 0.0021 & 0.0034 & 0.0030 & 0.0033 & 0.0042 & 0.0062 & 0.011 & \kin{0.024} & \kin{0.072} & \kin{0.44} & \kin{36.1} & \kin{$3.5\!\times\!10^{4}$} & \kin{$9.0\!\times\!10^{6}$} \\[4pt] \hline
\multirow{4}{*}{1 TeV}
 & $W^+W^-$         & 0.056 & 0.081 & 0.070 & 0.071 & 0.085 & 0.12 & 0.18 & 0.35 & \kin{0.86} & \kin{3.3} & \kin{56.3} & \kin{$1.4\!\times\!10^{5}$} & \kin{$2.7\!\times\!10^{7}$} \\
 & $\tau^+\tau^-$   & 0.033 & 0.047 & 0.041 & 0.042 & 0.050 & 0.067 & 0.11 & 0.20 & \kin{0.50} & \kin{2.0} & \kin{32.9} & \kin{$8.2\!\times\!10^{4}$} & \kin{$1.6\!\times\!10^{7}$} \\
 & $b\bar{b}$       & 0.65 & 0.94 & 0.81 & 0.83 & 0.99 & 1.3 & 2.1 & 4.0 & \kin{10.1} & \kin{39.0} & \kin{656} & \kin{$1.6\!\times\!10^{6}$} & \kin{$3.1\!\times\!10^{8}$} \\
 & $\nu\bar{\nu}$   & 0.0025 & 0.0037 & 0.0031 & 0.0032 & 0.0039 & 0.0052 & 0.0082 & 0.016 & \kin{0.039} & \kin{0.15} & \kin{2.5} & \kin{6373} & \kin{$1.2\!\times\!10^{6}$} \\[4pt] \hline
\multirow{4}{*}{4 TeV}
 & $W^+W^-$         & 0.75 & 1.0 & 0.87 & 0.87 & 1.0 & 1.3 & 2.0 & 3.5 & 8.2 & \kin{28.3} & \kin{350} & \kin{$1.3\!\times\!10^{6}$} & \kin{$2.0\!\times\!10^{8}$} \\
 & $\tau^+\tau^-$   & 0.24 & 0.33 & 0.28 & 0.28 & 0.32 & 0.42 & 0.62 & 1.1 & 2.6 & \kin{9.0} & \kin{111} & \kin{$4.3\!\times\!10^{5}$} & \kin{$6.3\!\times\!10^{7}$} \\
 & $b\bar{b}$       & 1.3 & 1.8 & 1.5 & 1.5 & 1.7 & 2.2 & 3.3 & 6.0 & 13.9 & \kin{48.4} & \kin{597} & \kin{$2.3\!\times\!10^{6}$} & \kin{$3.4\!\times\!10^{8}$} \\
 & $\nu\bar{\nu}$   & 0.048 & 0.066 & 0.056 & 0.056 & 0.064 & 0.084 & 0.13 & 0.23 & 0.52 & \kin{1.8} & \kin{22.4} & \kin{$8.6\!\times\!10^{4}$} & \kin{$1.3\!\times\!10^{7}$} \\[4pt] \hline\hline
\end{tabular}%
}
\vspace{0.2cm}
\caption{As in Table~\ref{tab:IceCube_O1s}, but for the spin-independent isovector $\mathcal{O}_1^v$ interaction. We also include the elastic limit, for which LZ reports a nonzero local significance~\cite{LZ:2026axp}.}
\label{tab:IceCube_O1v}
\end{table*}



\begin{table*}[tp!]
\centering
\renewcommand{\arraystretch}{1.4}
\resizebox{\textwidth}{!}{%
\begin{tabular}{c|c|cccccccccc}
\hline\hline
\multirow{3}{*}{$m_\chi$} & & \multicolumn{10}{c}{IceCube 90\% CL limits on $\mathcal{O}_{4}^{s}$ $\sigma_{\rm SD}$ $[\times 10^{-40}\,\text{cm}^{2}]$} \\[2pt] \cline{3-12}
 & \rule{0pt}{12pt} SM channel & \multicolumn{10}{c}{Mass Splitting $\delta$ [keV]} \\[2pt]
 & & 0 & 50 & 100 & 150 & 200 & 250 & 300 & 350 & 400 & 450 \\ \hline
\multirow{3}{*}{400 GeV}
 & $W^{+}W^{-}$     & 0.071 & 20.7 & 30.9 & 65.4 & 347 & 3473 & $2.0\!\times\!10^{4}$ & \kin{$2.0\!\times\!10^{5}$} & \kin{$4.6\!\times\!10^{6}$} & \kin{$4.6\!\times\!10^{8}$} \\
 & $b\bar{b}$       & 3.1 & 915 & 1365 & 2883 & $1.5\!\times\!10^{4}$ & $1.5\!\times\!10^{5}$ & $9.0\!\times\!10^{5}$ & \kin{$8.8\!\times\!10^{6}$} & \kin{$2.0\!\times\!10^{8}$} & \kin{$2.0\!\times\!10^{10}$} \\
 & $\nu\bar{\nu}$ (1:1:1) & 0.0029 & 0.86 & 1.3 & 2.7 & 14.3 & 143 & 845 & \kin{8212} & \kin{$1.9\!\times\!10^{5}$} & \kin{$1.9\!\times\!10^{7}$} \\[4pt] \hline
\multirow{3}{*}{1 TeV}
 & $W^{+}W^{-}$     & 0.054 & 12.7 & 17.9 & 33.9 & 154 & 1486 & 7432 & $5.4\!\times\!10^{4}$ & \kin{$8.3\!\times\!10^{5}$} & \kin{$4.0\!\times\!10^{7}$} \\
 & $b\bar{b}$       & 1.1 & 267 & 377 & 713 & 3240 & $3.1\!\times\!10^{4}$ & $1.6\!\times\!10^{5}$ & $1.1\!\times\!10^{6}$ & \kin{$1.7\!\times\!10^{7}$} & \kin{$8.3\!\times\!10^{8}$} \\
 & $\nu\bar{\nu}$ (1:1:1) & 0.0048 & 1.1 & 1.6 & 3.0 & 13.7 & 132 & 662 & 4851 & \kin{$7.4\!\times\!10^{4}$} & \kin{$3.5\!\times\!10^{6}$} \\[4pt] \hline
\multirow{3}{*}{4 TeV}
 & $W^{+}W^{-}$     & 0.44 & 103 & 145 & 274 & 1248 & $1.2\!\times\!10^{4}$ & $5.8\!\times\!10^{4}$ & $3.8\!\times\!10^{5}$ & $5.0\!\times\!10^{6}$ & \kin{$1.9\!\times\!10^{8}$} \\
 & $b\bar{b}$       & 2.4 & 567 & 798 & 1503 & 6856 & $6.8\!\times\!10^{4}$ & $3.2\!\times\!10^{5}$ & $2.1\!\times\!10^{6}$ & $2.7\!\times\!10^{7}$ & \kin{$1.1\!\times\!10^{9}$} \\
 & $\nu\bar{\nu}$ (1:1:1) & 0.082 & 19.3 & 27.2 & 51.3 & 234 & 2313 & $1.1\!\times\!10^{4}$ & $7.1\!\times\!10^{4}$ & $9.3\!\times\!10^{5}$ & \kin{$3.6\!\times\!10^{7}$} \\[4pt] \hline\hline
\end{tabular}%
}
\caption{As in Tables~\ref{tab:IceCube_O1s}-\ref{tab:IceCube_O1v}, but for the case of the spin-dependent isoscalar $\mathcal{O}_4^s$ interaction. We extend the scan up to $\delta=450$~keV, above which the inferred cross sections become large enough that the single-scatter treatment adopted here may not be reliable.}
\vspace{2.2cm}
\label{tab:O4s}
\end{table*}

\vspace{1cm}

\begin{table*}[tp!]
\centering
\renewcommand{\arraystretch}{1.4}
\resizebox{\textwidth}{!}{%
\begin{tabular}{c|c|cccccccccc}
\hline\hline
\multirow{3}{*}{$m_\chi$} & & \multicolumn{10}{c}{IceCube 90\% CL limits on $\mathcal{O}_{4}^{v}$ $\sigma_{\rm SD}$ $[\times 10^{-40}\,\text{cm}^{2}]$} \\[2pt] \cline{3-12}
 & \rule{0pt}{12pt} SM channel & \multicolumn{10}{c}{Mass Splitting $\delta$ [keV]} \\[2pt]
 & & 0 & 50 & 100 & 150 & 200 & 250 & 300 & 350 & 400 & 450 \\ \hline
\multirow{3}{*}{400 GeV}
 & $W^{+}W^{-}$     & 0.071 & 24.0 & 37.7 & 81.2 & 416 & 3520 & $2.0\!\times\!10^{4}$ & \kin{$2.0\!\times\!10^{5}$} & \kin{$4.6\!\times\!10^{6}$} & \kin{$4.6\!\times\!10^{8}$} \\
 & $b\bar{b}$       & 3.1 & 1059 & 1663 & 3581 & $1.8\!\times\!10^{4}$ & $1.6\!\times\!10^{5}$ & $9.0\!\times\!10^{5}$ & \kin{$8.8\!\times\!10^{6}$} & \kin{$2.0\!\times\!10^{8}$} & \kin{$2.0\!\times\!10^{10}$} \\
 & $\nu\bar{\nu}$ (1:1:1) & 0.0029 & 0.99 & 1.6 & 3.4 & 17.2 & 145 & 845 & \kin{8212} & \kin{$1.9\!\times\!10^{5}$} & \kin{$1.9\!\times\!10^{7}$} \\[4pt] \hline
\multirow{3}{*}{1 TeV}
 & $W^{+}W^{-}$     & 0.054 & 14.8 & 21.9 & 42.3 & 188 & 1516 & 7432 & $5.4\!\times\!10^{4}$ & \kin{$8.3\!\times\!10^{5}$} & \kin{$4.0\!\times\!10^{7}$} \\
 & $b\bar{b}$       & 1.1 & 311 & 461 & 890 & 3958 & $3.2\!\times\!10^{4}$ & $1.6\!\times\!10^{5}$ & $1.1\!\times\!10^{6}$ & \kin{$1.7\!\times\!10^{7}$} & \kin{$8.3\!\times\!10^{8}$} \\
 & $\nu\bar{\nu}$ (1:1:1) & 0.0048 & 1.3 & 2.0 & 3.8 & 16.8 & 135 & 662 & 4851 & \kin{$7.4\!\times\!10^{4}$} & \kin{$3.5\!\times\!10^{6}$} \\[4pt] \hline
\multirow{3}{*}{4 TeV}
 & $W^{+}W^{-}$     & 0.44 & 121 & 181 & 347 & 1545 & $1.3\!\times\!10^{4}$ & $5.8\!\times\!10^{4}$ & $3.8\!\times\!10^{5}$ & $5.0\!\times\!10^{6}$ & \kin{$1.9\!\times\!10^{8}$} \\
 & $b\bar{b}$       & 2.4 & 667 & 993 & 1906 & 8488 & $6.9\!\times\!10^{4}$ & $3.2\!\times\!10^{5}$ & $2.1\!\times\!10^{6}$ & $2.7\!\times\!10^{7}$ & \kin{$1.1\!\times\!10^{9}$} \\
 & $\nu\bar{\nu}$ (1:1:1) & 0.082 & 22.8 & 33.9 & 65.0 & 290 & 2365 & $1.1\!\times\!10^{4}$ & $7.1\!\times\!10^{4}$ & $9.3\!\times\!10^{5}$ & \kin{$3.6\!\times\!10^{7}$} \\[4pt] \hline\hline
\end{tabular}%
}
\caption{As in Tables~\ref{tab:IceCube_O1s}-\ref{tab:O4s}, but for the spin-dependent isovector $\mathcal{O}_4^v$ interaction. Comparing our results to Table~\ref{tab:O4s}, the limits differ by at most a few tens of percent for the tabulated points and converge at large splittings in the adopted nuclear-spin approximation. We similarly end our scan at a mass splitting of $\delta$~=~450~keV, above which the single-scatter approximation may not apply.}
\label{tab:O4v}
\vspace{0.1cm}

\end{table*}

\begin{figure*}[t!]
\centering
\includegraphics[width=1.95\columnwidth]{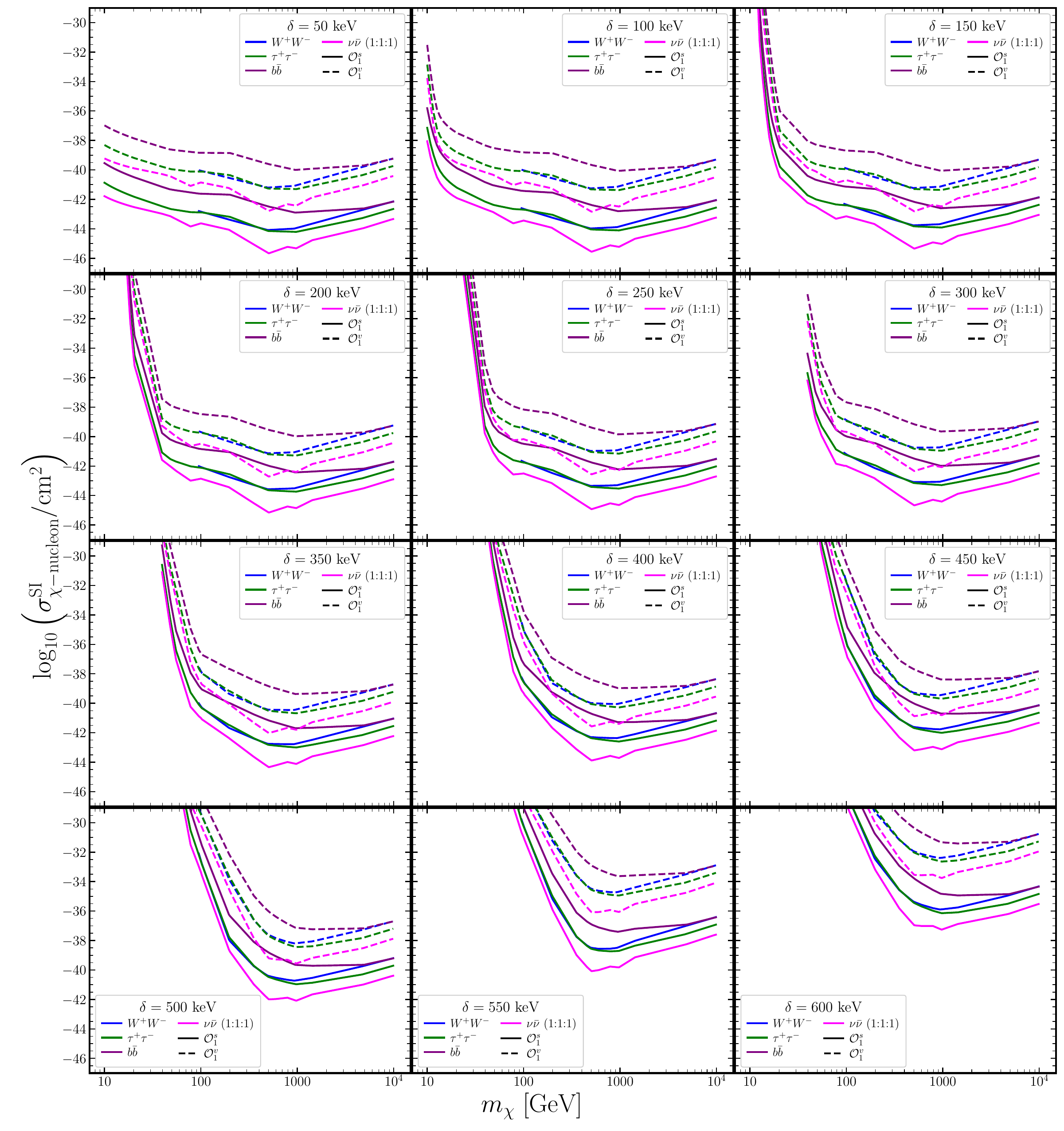}
\caption{IceCube constraints on the spin-independent inelastic DM-nucleon cross section for different mass splittings, DM masses, and annihilation channels.}
\label{fig:Inelastic_both}
\end{figure*}

\begin{figure*}[t!]
\centering
\includegraphics[width=2\columnwidth]{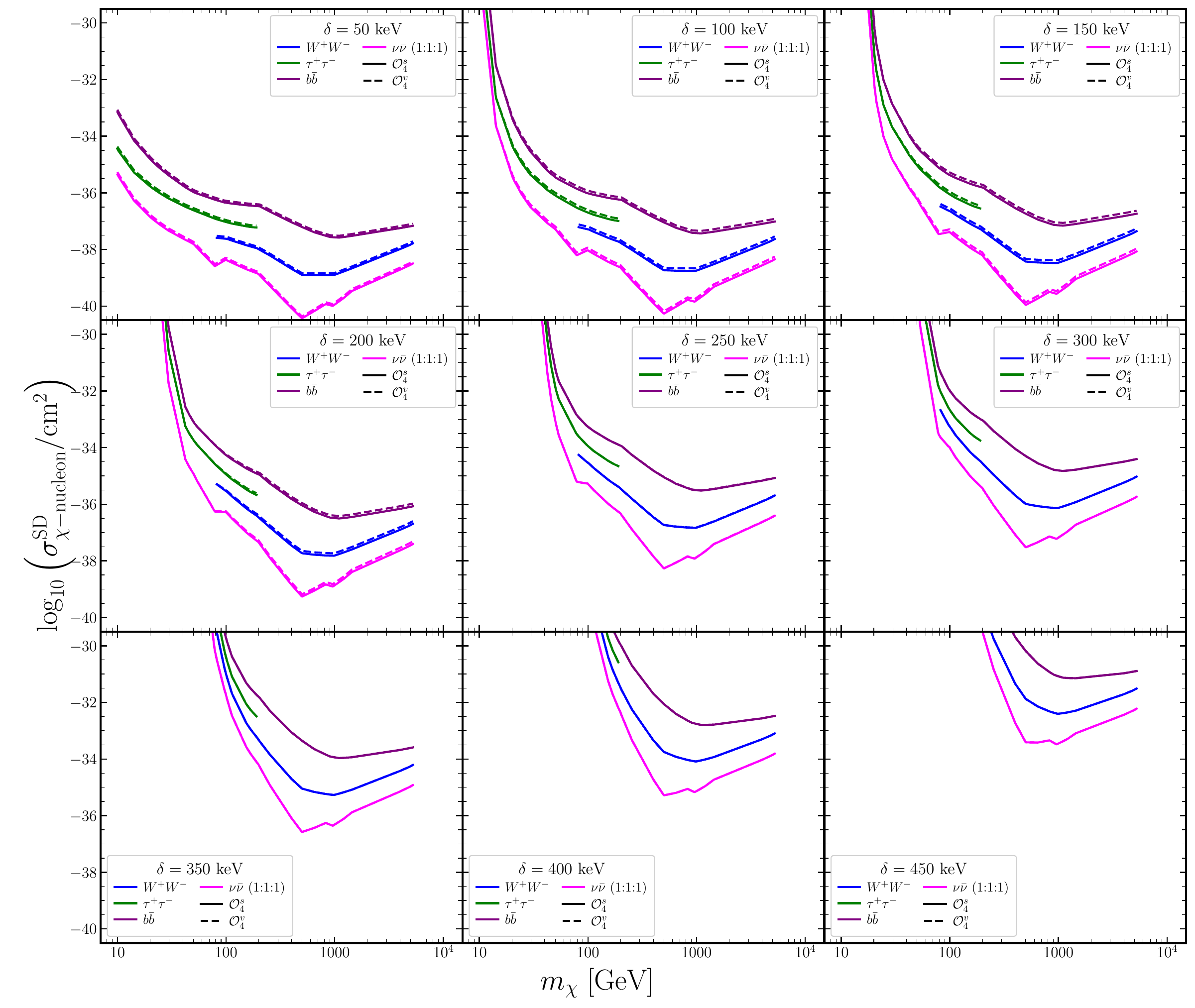}
\caption{IceCube constraints on the spin-dependent inelastic DM-proton cross section for different mass splittings, DM masses, and annihilation channels.}
\label{fig:Inelastic_O4}
\end{figure*}

If a sufficiently high density of DM accumulates in the solar core, these particles can annihilate to produce an observable high-energy neutrino flux~\cite{Srednicki:1986vj, Berlin:2024lwe, Maity:2023rez, Nguyen:2025ygc, Krishna:2025ncv, Hooper:2025ohk, Nguyen:2026apa}. While most DM annihilation products are absorbed by solar material, neutrinos produced directly or through the decays of short-lived particles can propagate through the Sun and reach Earth. For this reason, IceCube's solar observations are sensitive to DM annihilating directly to $\nu\bar{\nu}$, as well as to combinations of $W^+W^-$, $ZZ$, $hh$, $hZ$, $\tau^+\tau^-$, $b\bar{b}$, $c\bar{c}$, or $t\bar{t}$~\cite{Baratella:2013fya, Liu:2020ckq}.

The DM annihilation rate in the Sun is $\Gamma_{\chi\chi}=\tfrac12 A_\odot N_\chi^2$, where $N_{\chi}$ is the number of DM particles in the Sun and the annihilation rate per particle pair is given by
\begin{equation}
    A_{\odot} = \langle \sigma_{\chi \chi} v \rangle \frac{\int_{0}^{R_{\odot}}{\rm d}r\, 4\pi r^{2}n_{\chi}^{2}(r)}{\left( \int_{0}^{R_{\odot}} {\rm d}r\, 4\pi r^{2} n_{\chi}(r) \right)^{2}}.
\end{equation}

In evaluating this rate, we adopt an $s$-wave annihilation cross section of $\langle\sigma_{\chi\chi}v\rangle=3\times10^{-26}$~cm$^3$/s for the $\mathcal{O}_1^s$, $\mathcal{O}_1^v$, $\mathcal{O}_4^s$, and $\mathcal{O}_4^v$ benchmarks. For higgsinos, we adopt the annihilation cross section from Ref.~\cite{Dessert:2022evk}. The dominant higgsino annihilation channels are $W^+W^-$ and $ZZ$. Because the IceCube limits used here do not include a separate $ZZ$ result, we approximate the neutrino yield from all higgsino annihilations by assuming a 100\% annihilation rate into W$^+$W$^-$. This approximation is relatively accurate due to the very similar neutrino yields of these two channels. For the thermalized DM density profile, we follow Ref.~\cite{Garani:2017jcj}.

As the DM density builds, the $\rho^2$ dependence of the DM annihilation rate means that capture and annihilation eventually reach equilibrium. The timescale for the DM annihilation rate to reach equilibrium with its capture rate is given by
\begin{equation}
    t_{\rm eq} = \frac{1}{\sqrt{C_{\odot}A_{\odot}}}.
\end{equation}
Neglecting evaporation (which is not relevant for the DM masses considered in this study) and assuming constant capture and annihilation coefficients and an initially negligible DM population, the total annihilation rate is
\begin{equation}
    \Gamma_{\chi \chi}=\frac{C_{\odot}}{2}\tanh^{2}\left( \frac{t_{\odot}}{t_{\rm eq}} \right),
\end{equation}
where $t_{\odot}\approx 4.6$~Gyr is the solar age. This follows from $\dot N_\chi=C_\odot-A_\odot N_\chi^2$ and approaches $C_\odot/2$ when $t_\odot\gg t_{\rm eq}$.

\section{Results}
\label{sec:result}

\vspace{-0.3cm}

The IceCube Collaboration has reported constraints on DM annihilation in the Sun proceeding to $W^{+}W^{-}$, $\tau^{+}\tau^{-}$, $b\bar{b}$, and $\nu \bar{\nu}$~\cite{IceCube:2021xzo, IceCube:2025fcu}. Here, we recast these limits for inelastic DM-nucleus scattering. In Fig.~\ref{fig:O1s_constraints}, we show constraints for spin-independent, inelastic scattering with $m_{\chi} = 1 \, {\rm TeV}$ and isoscalar couplings ($f_p=f_n$). The gray region represents the parameter space that could account for the event reported by the LZ Collaboration. We find that that the majority of this parameter space is in tension with IceCube data. Note that the parameter space with $\delta \lsim 100 \, {\rm keV}$ cannot account for the L230616 event without predicting an unacceptable rate of low-energy nuclear recoils, while smaller mass splittings that lie near this kinematic limit (and are more consistent with our constraints) produce significantly poorer fits to the LZ data than larger mass splittings~\cite{LZ:2026axp}.

In Fig.~\ref{fig:O1_extend} we show the isoscalar (left) and isovector (right) results at $m_\chi=1$~TeV, and extend our calculation to $\delta=600$~keV. At our $m_{\chi}=1 \, {\rm TeV}$ benchmark mass, we find that solar limits generally remain strong up to mass splittings of $\sim$550~keV, above which the collisions of 1~TeV DM particles with even the heaviest solar nuclei no longer exceed the kinematic cutoff for the inelastic interactions. Notably, in the case of solar interactions, this conclusion depends primarily on the solar gravitational potential, and is only negligibly affected by uncertainties in the galactic DM halo velocity. These results indicate that IceCube constraints robustly rule out much of the parameter space where inelastic DM accounts for the LZ event, so long as a large fraction of DM annihilations produce neutrinos through a combination of direct neutrino annihilation, or annihilations to heavy fermions, and/or gauge or Higgs bosons.

One potential method of avoiding these constraints is to consider only spin-dependent DM interactions. Spin-dependent capture weakens sharply as the splitting increases from zero due to the fact that Hydrogen dominates the elastic limit in the spin-dependent case, but its small mass limits the inelastic threshold to
\begin{equation}
 \delta < \frac12\mu_{\chi p}w^2
 \approx 10~\mathrm{keV} \, 
 \left(\frac{w}{1400~\mathrm{km/s}}\right)^2.
\end{equation}
Thus, contributions from solar hydrogen already become inaccessible at splittings of order 10~keV. Solar capture must then proceed based only on the much less abundant spin-carrying isotopes, such as $^{14}$N, $^{23}$Na, $^{27}$Al, and heavier nuclei. Notably, LZ considers two realizations of spin-dependent DM models, the ($\mathcal{O}^s_4$, $f_p=f_n$) and isovector ($\mathcal{O}^v_4$, $f_p=-f_n$) interactions, finding that they fit the data with similar statistical significances to spin-independent operators. However, they do not directly calculate the spin-dependent cross-sections necessary to fit their data. 

In Fig.~\ref{fig:O4_extend}, we quantitatively consider this scenario. We definite the reference cross-section for both our $\mathcal{O}^s_4$ and $\mathcal{O}^v_4$ operators for protons, but include capture on all spin-carrying target isotopes. For the $m_\chi=1$~TeV benchmark, we omit $\tau^+\tau^-$ because the IceCube dataset used for this spin-dependent comparison reports that channel only for DM masses below 200~GeV~\cite{IceCube:2025fcu}. As expected, the kinematic loss of the dominant hydrogen targets at $\sim$10~keV produces a nearly factor of 100 weakening in the IceCube limit. However, the contribution from heavier nuclei is non-negligible, and we find that IceCube observations can still set relatively strong limits up to mass splittings of several hundred keV. Over most of the inelastic mass range, we obtain cross-section constraints which are roughly six orders of magnitude weaker than in the spin-independent case, which is relatively similar to the weakened cross-sections typically found when comparing elastic spin-dependent and spin-independent cross-sections. While we are unable to compare these results directly to LZ data, we find that solar observations are capable of sensitively probing this class of models. 

Thus far, we have only considered DM models at a benchmark mass of 1~TeV. In Tables~\ref{tab:IceCube_O1s}-\ref{tab:O4v}, we generalize our results to other DM masses. We provide constraints for three values of the DM mass, and for a range of mass splittings, interactions, and annihilation channels. In Figs.~\ref{fig:Inelastic_both} and~\ref{fig:Inelastic_O4}, we plot our constraints on the spin-independent and spin-dependent inelastic DM-nucleon cross section as a function of the DM mass, assuming a number of mass splittings and annihilation final states. These results indicate that solar neutrino constraints can sensitively probe inelastic DM interactions up to masses of $\sim$10~TeV, and up to inelastic mass splittings of $\sim$500~keV over a wide variety of assumptions regarding the specific DM/nucleon cross-section and annihilation final state, providing a complementary probe of high-energy nuclear recoil events in terrestrial detectors.

It is worth mentioning several features of our results. We find that solar constraints are strongest for DM masses between 500--1000~GeV. The sensitivity decreases at low DM masses due to a combination of the 10~GeV threshold of the IceCube neutrino analysis and the kinematics of the inelastic scattering cross-section. At high DM masses, the sensitivity decreases due to the smaller DM number density and the fact that the Sun becomes more thick to neutrinos. Throughout this mass range, direct annihilation to neutrinos produce a constraint that is 10--100$\times$ stronger than annihilations to $\tau^+\tau^-$ or W$^+$W$^-$, while $b\bar{b}$ constraints are noticeably weaker.

In the case of spin-dependent inelastic DM scattering, comparing Tabs.~\ref{tab:O4s}~and~\ref{tab:O4v} shows that the isoscalar and isovector limits converge above $\delta \approx 250$~keV. Larger splittings select heavier targets because $\delta_{\max}=\tfrac12\mu_{\chi A}w_{\max}^2$. In our spin-dependent calculations, $^{57}$Fe supplies 82\% of the capture rate at $\delta=250$~keV and 94\% at 450~keV. With the adopted matrix elements, $\langle S_n\rangle\simeq0.49$ and $\langle S_p\rangle\simeq0$, changing the sign of $a_n/a_p$ leaves $|\langle S_p\rangle+(a_n/a_p)\langle S_n\rangle|^2$ unchanged. An analogous argument applies to targets whose spin is assigned to protons. The convergence stems from the nuclear-spin approximation; it is not a general identity of the operators.

\begin{figure}[t!]
\centering
\includegraphics[width=1\columnwidth]{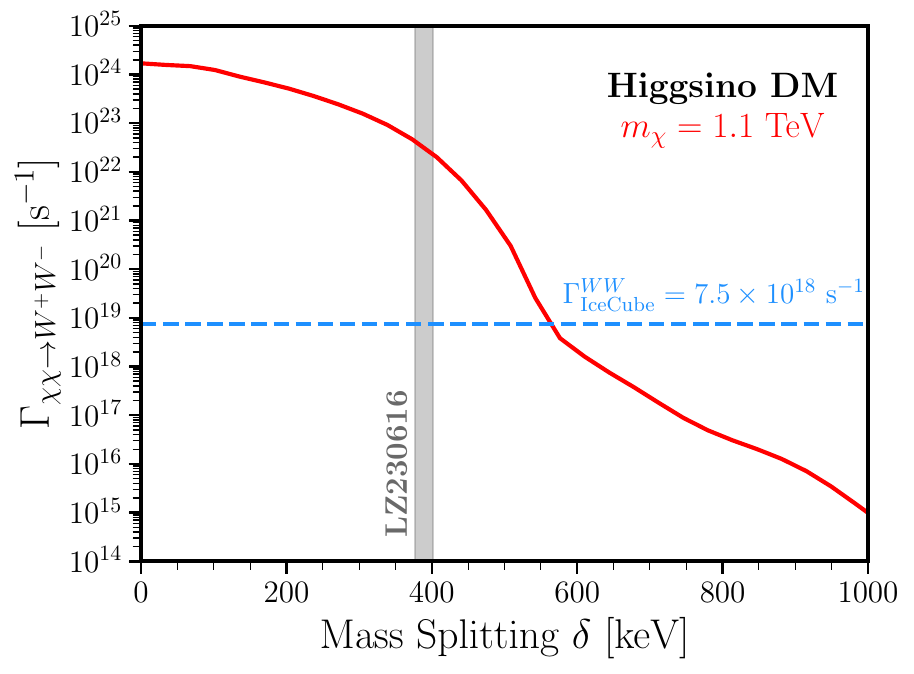}
\vspace*{-0.4cm}
\caption{The Solar DM annihilation rate for thermal higgsino DM (red), compared with the IceCube 90\% CL upper limit obtained using the $W^+W^-$ neutrino yield (light blue). The grey band denotes the mass-splitting interval favored by the LZ event. Our results are consistent with Refs.~\cite{Pospelov:2026ewn, Bose:2026ndd, DiMauro:2026dqp}.}
\vspace*{-0.4cm}
\label{fig:Higgsino}
\end{figure}

However, at small splittings, the limits for spin-dependent scattering differ by 15-30\%. Nonzero proton and neutron contributions allow interference, as in the adopted matrix elements for $^{23}$Na and $^{27}$Al. More complete nuclear calculations can give nonzero contributions from both nucleon species even when a simple approximation assigns the spin to only one. Thus, the degeneracy observed for large mass-splittings should be understood as a feature of the adopted matrix elements. This differs from spin-independent scattering, where the coherent nuclear factor is $[Z+(f_n/f_p)(A-Z)]^2$ and isovector cancellation can strongly suppress capture.

\begin{figure*}[t!]
\centering
\includegraphics[width=2\columnwidth]{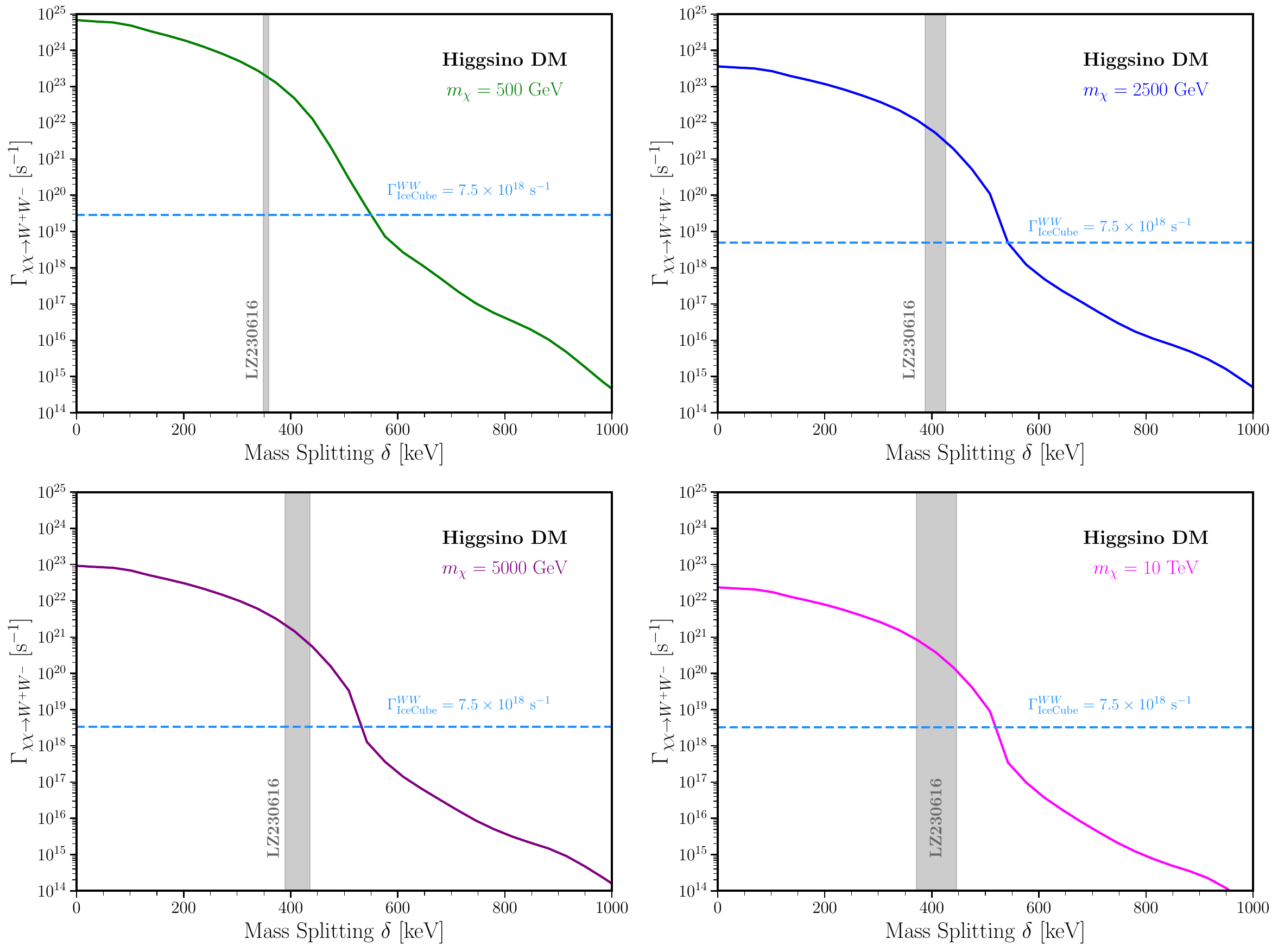}
\caption{As in Fig.~\ref{fig:Higgsino}, but for four nonthermal higgsino benchmark models from Ref.~\cite{Rodd:2026tyn}.}
\label{fig:Higgsino_nonthermal}
\end{figure*}

\subsection{Higgsino Dark Matter}
\label{ssec:Higgsino}
\vspace{-0.3cm}

Our methodology and results in this section are similar to those of Pospelov~and~Ramani~\cite{Pospelov:2026ewn}. We calculate the solar capture and annihilation rates of higgsino DM and compare the predicted annihilation rate with the IceCube upper limit, using the $W^+W^-$ channel as a proxy for the higgsino neutrino yield.

Fig.~\ref{fig:Higgsino} shows the predicted solar annihilation rate for a thermal higgsino with $m_{\chi} = 1.1 \, {\rm TeV}$ as a function of the mass splitting. We also show the annihilation rate limits from IceCube~\cite{IceCube:2025fcu}. For the adopted LZ-compatible interval, $\delta=377$-402~keV, our predicted rate exceeds this limit by approximately four orders of magnitude. This supports the conclusion of Pospelov and Ramani~\cite{Pospelov:2026ewn} that IceCube's observations strongly disfavor the thermal higgsino interpretation of the LZ event, at least under conventional halo assumptions.

We also confront the possibility that a nonthermal higgsino could explain the LZ event~\cite{Fan:2026kxx, Rodd:2026tyn}. For the four mass benchmarks from Ref.~\cite{Rodd:2026tyn}, we calculate the solar annihilation rates and compare them with the corresponding IceCube limits. Fig.~\ref{fig:Higgsino_nonthermal} shows these comparisons and the adopted LZ-compatible splitting intervals. These results demonstrate that IceCube excludes the tested benchmarks up to a mass of 10~TeV. While this work was being completed, Ref.~\cite{Bose:2026ndd} independently studied thermal and nonthermal higgsinos and also found substantial tension with solar-neutrino observations.

\subsection{Open Windows for Inelastic DM Interpretations of the Recent LZ Event}
\vspace{-0.3cm}

Although high-energy solar neutrino observations strongly constrain many of the inelastic DM models considered above, these limits rely on the efficient capture and large annihilation rate into neutrino-producing final states. Several possibilities could be invoked to weaken or evade these constraints.

\begin{itemize}
    
\item First, DM annihilation could be suppressed by the presence of a particle-antiparticle asymmetry~\cite{Petraki:2013wwa, Murase:2016nwx, Chu:2024gpe}. Celestial probes, such as neutron stars, can be sensitive to this scenario~\cite{Bramante:2014zca, Baryakhtar:2017dbj, Bramante:2017ulk, Saha:2025fgu, Bhattacharya:2024pmp, Bramante:2023djs}, but are typically less sensitive than direct detection in scenarios in which the DM particle is a fermion~\cite{Robles:2025dlv}.

\item Second, the DM could annihilate predominantly into dark-sector mediators whose decays produce few detectable neutrinos, or whose lifetimes and boost factors allow them to travel well past the Earth before decaying~\cite{Pospelov:2026ewn, deLima:2026shq, Zhu_2026_DP}. Recent work~\cite{deLima:2026shq, Zhu_2026_DP} has suggested that such models could explain the LZ event. In addition to neutrinos, however, models with escaping mediators can often decay to generate observable gamma rays, or charged cosmic rays, which often provide even stronger constraints unless the decay length is much longer than 1~AU~\cite{Linden:2024uph, Leane:2017vag, Smolinsky:2017fvb, Feng:2016ijc, Bell:2021pyy, Leane:2021ihh, Acevedo:2024ttq, Nguyen:2022zwb,Leane:2024bvh}.

\item Third, if only $\chi_1\chi_2$ coannihilation is allowed, the annihilation rate will depend on the abundance of both states~\cite{Berlin:2025fwx, Baker:2015qna}. This rate could thus be suppressed if one of the captured populations remains much smaller than the other.

\item Fourth, neutron-dominated spin-dependent couplings can reduce the solar capture rate relative to spin-dependent interactions with a substantial proton coupling, especially if hydrogen would otherwise dominate~\cite{Nguyen:2026nhe}. Notably, xenon's spin-carrying isotopes, $^{129}$Xe and $^{131}$Xe, remain sensitive to neutron couplings, while for mass splittings of several hundred keV, however, the dominant solar target of hydrogen is already kinematically inaccessible. However, odd-neutron solar isotopes can still power a reasonable capture rate, as illustrated above by $^{57}$Fe. A neutron-only coupling therefore does not automatically evade the inelastic solar bounds.

\item Fifth, inefficient cooling could leave captured DM particles in extended orbits for long periods of time, reducing the resulting annihilation rate. Such scenarios may be particularly relevant if the loop-level elastic scattering cross-sections that are expected in most inelastic DM models are significantly inhibited. Likewise, capture-annihilation equilibrium could be delayed or evaded if the DM annihilation cross section lies far below the expected thermal value at low-velocities, as may be relevant, \emph{e.g.,} in $p$-wave annihilation models.

\item Sixth, if DM annihilations generate mostly first or second generation fermions (other than neutrinos), these particles will lose most of their energy in the solar medium before decaying to produce neutrinos. 

\end{itemize}

More generally, other interpretations of the LZ event could evade the constraints presented here if they do not involve conventional nuclear scattering followed by DM annihilation to an appreciable flux of high-energy neutrinos~\cite{Unwin:2026rdp, Visinelli:2026kgt, Smirnov:2026aqk, Wang:2026ytg, Kotlarski:2026pep, Lee:2026wof, Lou:2026idn, McCabe:2026crm}. 

\section{Conclusion and Outlook}
\label{sec:conclusion}

In this article, we have shown that IceCube limits on high-energy neutrinos from the Sun constrain inelastic dark matter-nucleus scattering over a broad range of masses and mass splittings. These results confirm the recent conclusions of Pospelov and Ramani for the case of pure-higgsino dark matter~\cite{Pospelov:2026ewn}, and extend the analysis to DM with a broad range of spin-independent and spin-dependent interactions.

The tables and figures included in this study provide solar-neutrino constraints which are complementary to the LZ analysis over a wide range of masses and mass splittings. These constraints can inform interpretations of the LZ event and future inelastic searches by XENONnT~\cite{XENON:2025vwd}, PandaX~\cite{PandaX-4T:2021bab}, and proposed paleodetectors~\cite{Graham:2026ivn}. More broadly, the interactions capable of producing a high-energy terrestrial recoil can also drive efficient solar capture. A detectable neutrino signal is not necessarily guaranteed, but the combination of direct detection and solar-neutrino searches can test both the scattering interaction and the subsequent evolution of captured dark matter.

We note that inelastic DM interpretations of the LZ event could remain viable if solar capture or annihilation is inefficient, or if the DM annihilations proceed to channels which result in few high-energy neutrinos. 

\vspace*{0.1cm}

\section*{Acknowledgements}
\vspace*{-0.3cm}
TTQN and TL are supported by the Swedish Research Council under contract 2022-04283. TTQN is also supported by two grants from the Royal Swedish Academy of Sciences~(KVA): PH2025-0073 (Physics) and AST2025-0048 (Astronomy and Space Science). TL is also supported by the Swedish National Space Agency under contract 117/19. TTQN and TL acknowledge the support from EDUCATE Excellence Centre funded by the Swedish Research Council through grant Dnr~2022-06627. DH is supported by the Office of the Vice Chancellor for Research at the University of Wisconsin–Madison with funding from the Wisconsin Alumni Research Foundation. Parts of this work were performed using computing resources provided by the National Academic Infrastructure for Supercomputing in Sweden (NAISS) under Project 2025/5-729, which is partially funded by the Swedish Research Council through Grant 2022-06725.\\

\noindent {\bf AI Usage Statement.} Claude Code was used to assist with visually generating the plots as well as for grammar and typographical checks. AI tools were not used to generate scientific results, draw conclusions or to write any sections of the paper. All analysis, code, and the final text were produced, checked, and verified by the authors, who take full responsibility for the content of this work. 

\bibliography{main}

\end{document}